\documentclass[12pt]{article}
\usepackage{tikz}
\usepackage{pgfplots}
\usepackage[margin=1.25in]{geometry}
\usepackage[font=footnotesize,labelfont=bf]{caption}
\usepackage{longtable}
\usepackage{booktabs}
\usepackage{setspace}
\usepackage{graphicx}
\usepackage{float}
\usepackage{amsfonts}
\usepackage{amsmath}
\usepackage{natbib}

\usetikzlibrary{patterns}
\usetikzlibrary{shapes.geometric, arrows}
\pgfkeys{/pgf/number format/.cd,fixed,precision=3}
\pgfplotsset{
    compat=1.5.1, 
    every axis/.append style={scale only axis, axis on top,
        height=3.5cm, width=3.5cm, xmin=0, xmax=100, ymin=0, ymax=100, label style={font=\tiny}
        }
}
\pgfplotsset{every tick label/.append style={font=\tiny}}
\pgfplotsset{every axis title/.append style={font=\small}}
\tikzstyle{input} = [ellipse,minimum width=1cm,text centered,draw=black,line width=1pt]
\tikzstyle{hidden} = [circle,minimum width=0.5cm,text centered,draw=black,line width=1pt]
\tikzstyle{outnode} = [circle,minimum width=1.5cm,text centered,draw=black,line width=1pt]
\tikzstyle{arrow} = [->,>=stealth]
\tikzstyle{leftlabel} = [rectangle, align=right,draw=black!0]
\tikzstyle{rightlabel} = [rectangle, align=left,draw=black!0]

\begin{document}

\title{On the Equivalence of Neural and Production Networks}
\author{Roy Gernhardt \and Bjorn Persson}
\maketitle

\begin{abstract}
This paper identifies the mathematical equivalence
between economic networks of Cobb-Douglas agents and Artificial Neural
Networks. It explores two implications of this equivalence under general
conditions. First, a burgeoning literature has established that network
propagation can transform microeconomic perturbations into large aggregate
shocks. Neural network equivalence amplifies the magnitude and complexity of
this phenomenon. Second, if economic agents adjust their production and
utility functions in optimal response to local conditions, market pricing is
a sufficient and robust channel for information feedback leading to macro learning.
\end{abstract}

\section{Introduction}

There is a burgeoning literature surrounding the network origins of
aggregate shocks. Network economics can model how an individual's behavior
effects the society in which she is embedded \citep{Goyal}. Several
mathematical frameworks have been proposed that amplify microeconomic
fluctuations into large macroeconomic movements \citep{Acemoglu2012,
Acemoglu2015} which would otherwise be modeled as exogenous shocks. This
paper's core innovation is that it identifies a more natural and flexible mathematical model than previously examined. This model is intrinsic to existing theory. Specifically, economic networks of
Cobb-Douglas agents facing nonlinear inverse demand functions are
mathematically equivalent to Artificial Neural Networks (ANNs.)

We survey the pertinent mathematics at the intersection of the two
disciplines of Economics and Machine Learning to demonstrate this
equivalence. We explore two implications of modeling the economy as a Neural
Network (NN) under general conditions. The first is a consequence of the
functional completeness of NNs: neither the amplitude nor complexity of an
aggregate \textquotedblleft shock\textquotedblright\ in response to a given
microeconomic input can be reasonably bounded. Any conceivable functional
form is possible. Second, if the producers and consumers are free to adjust
their production and utility functions in optimal response to local
conditions, market pricing is a sufficient and robust channel for
information feedback leading to global, macro-scale learning. This learning is at the level of
the economy as a whole and is transparent to the individual
producers and consumers. 

Below, we first develop the mathematical equivalence between production
units and ANNs. Then we address the endogeneity/exogeneity of macro shocks,
and finally we discuss global learning behavior.

\section{Cobb-Douglas Producers and Artificial Neurons}

We posit that the Cobb-Douglas producers behave like Artificial Neural
Networks for two reasons. First, the microeconomic agents are mathematically
analogous to artificial neurons, the fundamental components of ANNs. Second,
the microeconomic agents interact with each other in a way that is
mathematically analogous to the interactions between neurons in an ANN. To
demonstrate this equivalence, we will begin with the ANN neuron.

ANN neurons combine weighted input signals, which are the outputs of other
neurons, in a nonpolynomial fashion. Theoretically, the inputs can be
weighted in any fashion (exponentially, polynomially, etc.), but in practice
what we call a Standard Artificial Neuron (SAN) takes the form of Equation \ref{SAN} %
\citep{Minsky}.

\begin{equation}\label{SAN}
P_{i}=a(b_{i}+\mathbf{\omega }_{i}\mathbf{\cdot p}_{k}),
\end{equation}

where $a(x)$ is an \textquotedblleft activation function\textquotedblright ; 
$\mathbf{p}_{k}$ is a vector of output values from the previous layer $k$; $%
\mathbf{\omega }_{i}$ is a vector of \textquotedblleft synaptic
weights\textquotedblright\ or coefficients belonging to neuron $i$; $b_{i}$
is a \textquotedblleft bias\textquotedblright\ belonging to neuron $i$; and $%
P_{i}$ is the output of neuron $i$. $P_{i}$ is destined to be one of the
entries in the vector $\mathbf{p}_{k+1}$ to be used in the next layer.

The activation function, $a(x)$, can take many forms, with different
functions having different properties, but as long as the activation
function is nonpolynomial an ANN with linear weights will function \citep{Cybenko, Ritter}.

Due to a number of desirable properties, one commonly used activation
function is the \textquotedblleft Softplus\textquotedblright\ function
expressed in Equation \ref{softplus} \citep{Glorot}:%
\begin{equation}\label{softplus}
\sigma \left( x\right) =\ln \left( 1+e^{x}\right)
\end{equation}%
We will modify this equation slightly:%
\begin{equation}\label{reflectedSoftplus}
a(x)=-\sigma (x)=\ln \left( \frac{1}{1+e^{x}}\right)
\end{equation}%
This will have the exact same properties as the Softplus, but it will
reflect the output around the x-axis. This inversion has no effect on the
network because the following layer's neurons can neutralize it by inverting
the sign of their weight vectors. The ANN neuron specification using the
inverted Softplus is equivalent (with the constraints we discuss below) to a
profit maximizing, price-taking Cobb-Douglas producer facing an unknown
inverse demand of the form\footnote{%
The inverse demand function is unknown to the producer in this model for two
reasons: first, it's a reasonable economic assumption that a price-taking
producer's price prediction is imperfect and second, it makes the math more
tractable. If we instead use the strong assumption that the function is
known to the producer, as is often done in simplified economic models, a
less elegant activation function is implied and no closed form profit
maximization solution exists for this particular inverse demand function.
This strong assumption would not affect NN equivalence because the price of
the producer's output would still be a nonpolynomial function of the prices
of her inputs. Therefore, our analysis still holds because the Economic Neural Network (ENN) is still
functionally complete (that is, any final output price function can be
approximated to an arbitrary degree of precision with correctly chosen
parameters and a large enough network.)}:%
\begin{equation}\label{inverseDemand}
\rho (Y_{i})=\frac{1}{1+Y_{i}}
\end{equation}%
Here $Y_{i}$ is the fully differentiated but substitutable output of
producer i and $\rho (Y_{i})$ yields the price of each unit of $Y_{i}$. This
inverse demand function is not mathematically necessary for economic agents
to form an ENN, but it has reasonable theoretical properties: It exhibits a
fixed maximum price, is decreasing in $Y$, and approaches zero
asymptotically.

Now we examine the Cobb-Douglas Producer-cum-neuron in the ENN. Let $m$ be
an index corresponding to a neuron in the previous layer; $A$ be the
\textquotedblleft technology parameter\textquotedblright ; $x_{ji}$ be the
input quantity demanded by producer $i$ for the jth good; $\alpha _{ji}$ be
the exponent in i's Cobb-Douglas function parameterizing the jth good; $P_{i}
$ be the price that producer i expects to get for each unit of her output at
the time of her production decision; $p_{j}$ be the price per unit of input
j.

The Decreasing Returns to Scale (DRS) Cobb-Douglas production function is:%
\begin{equation}\label{DRS_CD}
Y_{i}=A_{i}\prod_{m=1}^{n}x_{mi}^{\alpha _{mi}},
\end{equation}
where $x_{mi},\alpha _{mi}>0,\Phi _{i}<1,$ and where $\Phi _{i}\equiv \sum\limits_{m=1}^{n}\alpha _{mi}$ and $\eta _{i}\equiv \Phi _{i}-\alpha _{1i}.$ The profit function is:%
\begin{equation}\label{profit}
\pi _{i}=P_{i}Y_{i}-\sum_{m=1}^{n}p_{m}x_{mi}.
\end{equation}%
Maximizing profit, the first order conditions give us equations \ref{qtyFirstInput} and \ref{qtyMthInput}:%
\begin{equation}\label{qtyFirstInput}
x_{1,i}=\left[ \left( P_{i}A_{i}\right) ^{-1}\left( \frac{p_{1}}{\alpha _{1i}%
}\right) ^{1-\eta _{i}}\prod\limits_{m=2}^{n}\left( \frac{p_{m}}{\alpha
_{mi}}\right) ^{\alpha _{mi}}\right] ^{\frac{1}{\Phi _{i}-1}}.
\end{equation}

\begin{equation}\label{qtyMthInput}
x_{mi}=x_{1i}\frac{p_{1}\alpha _{mi}}{p_{m}\alpha _{1i}}
\end{equation}%
The ENN model requires output in terms of prices alone. Inserting equations
\ref{qtyFirstInput} and \ref{qtyMthInput} into equation \ref{DRS_CD} yields equation \ref{YfromPrices}:%
\begin{equation}\label{YfromPrices}
Y_{i}=\left[ P_{i}^{-\Phi _{i}}A_{i}^{-1}\prod\limits_{m=1}^{n}\left( \frac{%
p_{m}}{\alpha _{m,i}}\right) ^{\alpha _{m,i}}\right] ^{\frac{1}{\Phi _{i}-1}%
}.
\end{equation}%
Transforming \ref{YfromPrices} into logs yields equation \ref{logYfromPrices}:%
\begin{equation}\label{logYfromPrices}
\ln Y_{i}=\sum_{m=1}^{n}\frac{\alpha _{mi}}{\Phi _{i}-1}\ln p_{m}+\frac{\Phi
_{i}\ln P_{i}+\ln A_{i}+\sum_{m=1}^{n}\alpha _{mi}\ln \alpha _{mi}}{1-\Phi
_{i}}.
\end{equation}%
Now, let $\omega _{m}\equiv \frac{\alpha _{mi}}{\Phi _{i}-1}$, let $\omega $
be the vector of all $\omega _{m}$, let $l\equiv \ln p$ calculated
elementwise (that is, the vector of log input prices,) let 
\begin{equation*}
z_{i}=\frac{\Phi _{i}\ln P_{i}+\ln A_{i}+\sum_{m=1}^{n}\alpha _{mi}\ln
\alpha _{mi}}{1-\Phi _{i}},
\end{equation*}%
and let the log price of output, $L_{i}\equiv \ln \rho \left( Y_{i}\right)
=\ln \left( \frac{1}{1+Y_{i}}\right) $ by (7). Then $\ln Y_{i}=\mathbf{l}_{k}%
\mathbf{\cdot \omega }_{k}+z_{i}$ and $a\ln Y_{i}=L_{i}$ by (6), and the
price of the good output by a Cobb-Douglas producer can be modelled in logs
with equation \ref{victory}:%
\begin{equation}\label{victory}
L_{i}=a\left( \mathbf{l}_{k}\mathbf{\cdot \omega }_{k}+z_{i}\right) .
\end{equation}

This is the standard ANN neuron as expressed in \ref{SAN}. In the ENN context, the
market prices of the producers' inputs are equivalent to neuronal input
values and the price of the producer's output is equivalent to neuronal
output values. Although the mathematical model we show specifically invokes
the Cobb-Douglas producer, the equivalence between an economic agent
(whether goods producer or labor producing consumer) and an ANN neuron is
general and robust to production, utility, and inverse demand function
specifications. As long as the price of an output good cannot be expressed
as a linear combination of the prices of input goods, the production of that
output good within a network of similar production activities creates a
functionally complete network.

\section{Functional Completeness}

As we mentioned above, the Cobb-Douglas has constraints that do not bind the
ANN neuron.  The exponents of the
production function, $\alpha _{mi}$, must be greater than zero and the
function we specified must be DRS, that is, $\Phi _{i}$ must be less than
one (DRS is required so that the profit maximization process we use is
valid). Therefore, the weights $\omega _{mi}$ are strictly less than zero.
This limits the ENN's learning potential compared to an ANN which can have
both positive and negative weights. As our computer simulation shows below, this
constraint does not fully remove the ENN's ability to learn. Nor is this
constraint evident in real world production functions. It is an artifact of
the Cobb-Douglas, which we have chosen because of its prominence as a
theoretical standard rather than for empirical reasons.

Our choice of the sub-optimal Cobb-Douglas specification serves as both a
robustness test of ENN behavior and a nod to canonical economic theory. We
use it exclusively in our simulations for those reasons. A more realistic
ENN model, however, would include heterogenous household utility functions
and firm production functions that inherently overcome the limitations
imposed by homogenous Cobb-Douglas producers. In order to demonstrate how
easily a practical ENN specification can emulate the full range of ANN
behavior, we consider the NAND gate.

The NAND gate is functionally complete, and since a simple single hidden
layer NN can emulate the NAND gate, Neural Networks are also functionally
complete by extension. This means that all possible logical functions can be
approximated to an arbitrary degree of precision by an NN 
\citep{Abu-Mostafa,
Leshno, Ritter}. This argument is most intuitive when using neurons with
Heaviside step activation functions, but it is well understood in the field
that functional completeness extends naturally to Standard Artificial
Neurons with differentiable activation functions as specified above. The SAN
does not output two discreet values, but the values can be interpreted as
probabilities with an actionable probability threshold. If a neuron outputs
a value greater than or equal to the threshold, it can be interpreted as the
binary 1 (or electronic HIGH, or logical TRUE), and if below the threshold,
it can be interpreted as the binary 0 (or electronic LOW, or logical FALSE).%
\begin{equation*}
\begin{tabular}{|c|c|c|}
\hline
\multicolumn{3}{c}{\textbf{NAND Truth Table}} \\ \hline
Input 1 & Input 2 & Output \\ \hline
$0$ & $0$ & $1$ \\ \hline
$0$ & $1$ & $1$ \\ \hline
$1$ & $0$ & $1$ \\ \hline
$1$ & $1$ & $0$ \\ \hline
\end{tabular}%
\end{equation*}

Because the synaptic weights of the Cobb-Douglas Producer Neuron (CDPN) in
equation \ref{victory} are strictly negative, it cannot form a NAND gate. Neither can
it form a NOT gate, which could be used in series with an AND or OR gate to
build a functionally complete network. In order to establish the feasibility
of a functionally complete ENN, we will specify parameters of a CDPN to form
an AND gate and then specify a Cobb-Douglas-like production function and
parameters to form a NOT gate. The NOT gate can then be joined in series
with the AND to form the functionally complete NAND gate.

One example set of parameters with which a two-input CDPN can emulate an AND
gate is as follows (here using a sigmoid activation function: $a(x)=\ln
\left( \frac{1}{1+e^{x}}\right) ):$

\begin{center}
$\alpha _{1}=\alpha _{2}=\frac{20}{41}\qquad P_{i}=1\qquad A=\frac{e^{\frac{%
30}{41}}}{\alpha _{1}^{\Phi }}.$
\end{center}

Therefore, $\omega _{1}=\omega _{2}=-20,$ and $z=30.$ Let this CDPN be
called producer (or agent or neuron) $X.$ Let the producer emulating the NOT
gate be called the $\Psi $ neuron.

Let $\Psi $ have the capability of producing one of two products, $Y_{3}$ or 
$Y_{4}$, each using CD functions with a single input. But due to the
production technology employed, $\Psi $ cannot produce both $Y_{3}$ and $%
Y_{4}$. Optimal output of each product is calculated by $\Psi $, and she
then chooses to produce the product with the greater profit at that optimal
level, thereby forgoing any production of the less profitable product. This
production function satisfies the following expressions:%
\begin{equation}\label{atLeastOneOutputZero}
(Y_{3}=0)\vee \left( Y_{4}=0\right)
\end{equation}%
and

\begin{equation}\label{atLeastOneOutputPositive}
\left( Y_{j}=A_{j}x_{j}^{\alpha _{j}}\right) \vee \left( Y_{j}=0\right) 
\text{ for }j=3,4.
\end{equation}

Using the same process that transformed \ref{DRS_CD} into \ref{victory} above, \ref{atLeastOneOutputPositive} can be
re-expressed in terms of output prices, in logs:%
\begin{equation}\label{xformed_atLeastOneOutputPositive}
\ln \rho \left( Y_{j}\right) =a\ln Y_{j}=a(\omega _{j}\ln p_{j}+z_{j})\vee
1=L_{j}\text{ for }j=3,4.
\end{equation}

Producer $\Psi $, as expressed in (17) has two outputs, $L_{3}$ and $L_{4}$,
and two inputs, $p_{3}$ and $p_{4}.$If input $p_{4}$ is constant (or varies
within a range, or has less volatility than $p_{3}$) the parameters of $\Psi 
$ can be set so that its output $L_{4}$ and input $p_{3}$ form a NOT gate.

One set of parameters with which $\Psi $ can emulate a NOT gate (when $\ln
p_{4}\approx 0$) is as follows:

\begin{center}
$P_{3},$ $P_{4}=1,\alpha _{3}=\frac{1}{2}$, $A_{3}=3,$ $\alpha
_{4}=1-e^{-10},$ $A_{4}=\frac{1}{(1-\alpha _{4})^{1-\alpha _{4}}\alpha
_{4}^{\alpha _{4}}}.$
\end{center}

Combining agents $X$ and $\Psi $ with the parameters listed above into a
simple two neuron network yields a functioning NAND gate. Let the output of $%
X$ be $L_{3}\equiv \ln p_{3}$. That is, the output product of $X$ is the
first input of $\Psi $. The output of this network is shown below in
truth-table format. Since the NAND gate is functionally complete, any ENN
which incorporates producers who can shift production between two or more
products may also be functionally complete. Of course, other conceivable
production functions can yield the same results. The only requirement is
that some of the producers in the network experience, within some range of
conditions, circumstances in which an increase in the price of one or more
of their inputs induces an increase in the output of one or more of their
products.%
\begin{equation*}
\centering%
\begin{tabular}{|c|c|c|c|c|}
\hline
\multicolumn{5}{|c|}{\textbf{NAND gate ENN (with agents $X$ and }$\Psi $ )}
\\ \hline
lnp$_{1}$ & lnp$_{2}$ & Inp$_{3}$ & lnp$_{4}$ & L$_{4}$ \\ 
{\small (logical input 1)} & {\small (logical input 2)} & {\small ($\ln
p_{1}\wedge \ln p_{2})$} & {\small (assumed const.)} & {\small $\lnot $($\ln
p_{1}\wedge \ln p_{2})$} \\ \hline
$0$ & $0$ & 0 & 0 & 1 \\ \hline
$0$ & $1$ & 0 & 0 & 1 \\ \hline
$1$ & $0$ & 0 & 0 & 1 \\ \hline
$1$ & $1$ & 1 & 0 & 0 \\ \hline
\end{tabular}%
\ 
\end{equation*}

\section{\textbf{Complex Behavior in Simple Networks}}

Next we explore how small, local shocks can propagate through an ENN to generate
sometimes highly surprising aggregate behaviors. Here we build upon the
arguments of Vasco \citet{Carvalho2014}, who uses network structures to
\textquotedblleft confront a deep-seated and influential logic which, to
this day, justifies the continued appeal to an exogenous synchronization
device, in the form of aggregate shocks.\textquotedblright\ Our model
amplifies Carvalho's reasoning. When a laborer takes maternity or
bereavement leave or contracts a serious illness, or when a company changes
hands through inheritance, these micro shocks are not necessarily lost in
the random noise of the aggregate economy. Instead, they can become
important signals which reverberate through the markets and change the
direction of the economy as a whole.

A modern economy can be described as an intricate system of networks in
which consumers and producers interact via reasonably efficient markets.
Within these networks, small changes directly affecting only a small set of
producers within a sector of the economy may translate to large and
unforeseeable changes in the aggregate. This system is too complex to
describe in detail, and impossible to model explicitly using standard
economic models. Standard models often aggregate heterogeneity into
distributions according to various assumptions and thereby lose this
important feature of the economy they seek to model. Attempts have been
made, however, to better understand the inner workings of production
networks and how micro shocks in them produce macroeconomic fluctuations
(see e.g. \citet{Acemoglu2012}, \citet{Carvalho2014}).

Here we are considering a world where explicit modeling is not possible due
to the complexity of the various networks that comprise the economy. A key
feature is to retain the idiosyncrasies that characterize real economies and
how they propagate through the network, producing sometimes unpredictable,
inexplicable, and, not seldom, unintended consequences.

For tractability, we specifically consider a simple network architecture
with one input, one hidden layer, and a dual output. The set-up is inspired
by Rosenblatt's perceptron model in which binary classifiers learn to
recognize patterns through an algorithm when fed training data %
\citep{Rosenblatt}. Our model is a slightly modified version of the original
perceptron. The network is one directional so there are no feedback loops.
Consequently, unlike the perceptron model, learning cannot
occur. Output seems quite random, yet follows logical principles that are
unobservable. Outside empirical data can be fitted to the model, and this
data would be consistent with reality. But naturally, this data will be
overfitted. Standard models rely on exogenous variation (i.e., exogenous
shocks) to explain e.g. aggregate business cycle fluctuations -- here these
fluctuations arise naturally from individual changes in behavior that may
occur for endogenous reasons.

\begin{figure}[h]
\centering
\begin{tikzpicture}[node distance=2cm]

\node (rawmat)[input]{Raw Material Price};
\node (spacer)[leftlabel, below of=rawmat, yshift=.5cm]{};

\node (int1)[hidden, above of=rawmat, xshift=-3.2cm]{};
\node (int2)[hidden, above of=rawmat, xshift=-1.5cm]{};
\draw[line width=0pt, color=black!0](int1) -- node[color=black]{\huge...}(int2);
\node (leftlab)[leftlabel, left of=int1]{Intermediate Good\\Producers};

\draw[arrow,line width=1pt, color=black](rawmat) -- (int1);
\draw[arrow,line width=1pt, color=black](rawmat) -- (int2);

\node (leis1)[hidden, above of=rawmat, xshift=3.2cm]{};
\node (leis2)[hidden, above of=rawmat, xshift=1.5cm]{};
\draw[line width=0pt, color=black!0](leis1) -- node[color=black]{\huge...}(leis2);
\node (rightlab)[rightlabel, right of=leis1]{Leisure Good\\Producers};

\draw[arrow,line width=1pt, color=black](rawmat) -- (leis1);
\draw[arrow,line width=1pt, color=black](rawmat) -- (leis2);

\node (vendor)[outnode, above of=rawmat, yshift=2cm]{Vendor};
\node (hotdogs)[leftlabel, above of=vendor, xshift=-1cm, yshift=-.2cm]{Make\\Hotdogs};
\node (chillout)[rightlabel, above of=vendor, xshift=1cm, yshift=-.2cm]{Watch\\The Game};

\draw[arrow,line width=1pt, color=black](int1) -- (vendor);
\draw[arrow,line width=1pt, color=black](int2) -- (vendor);
\draw[arrow,line width=1pt, color=black](leis1) -- (vendor);
\draw[arrow,line width=1pt, color=black](leis2) -- (vendor);

\draw[arrow,line width=1pt, color=black](vendor) -- (hotdogs);
\draw[arrow,line width=1pt, color=black](vendor) -- (chillout);

\end{tikzpicture}
\captionsetup{font=footnotesize,labelfont=bf}
\caption{The four models in figure \protect\ref{funkyGraphs} are Cobb
Douglas producer networks. The so-called hidden layer has five producers in
models two and three and ten producers in models one and four. Hidden layer
members are either intermediate good producers or leisure good producers.
These goods are inputs to the network's final output producer. All hidden
layer production functions take the form $Y_i=A_i x^{\protect\alpha_i}$
where $x$ is the quantity of the sole raw material required to produce
intermediate and leisure goods in this economy, $Y_i$ is output, and $A_i$
is the technology parameter. The network's final producer outputs $%
Y_l=A_l\prod_{k=1}^K{{x_k}^{a_k}}\ $ where $l\in$ \{Intermediate, Leisure\}
and $k$ is the index of the $k$th labor or leisure good (as applicable). }
\label{networkDrawing}
\end{figure}
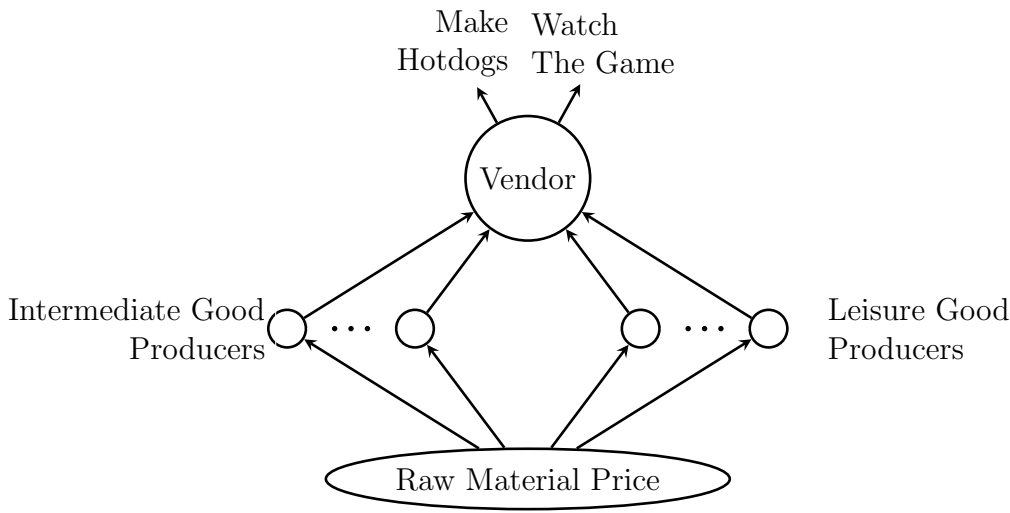

We use a basic Cobb-Douglas technology in which a single input price is fed
to a set of up to ten producers who produce intermediate goods to the final
producer in the economy. The final producer then chooses between two kinds
of outputs. We illustrate the ratio of these two outputs in Figure \ref%
{funkyGraphs} on page \pageref{funkyGraphs} with respect to the price of the
input commodity. In order for the impacts of these behaviors to be as clear
as possible, it is helpful to imagine the two products are Labor and Leisure.%
\footnote{%
There is a complicating factor which limits this computer model's accuracy
with regard to a producer's labor/leisure choice. It would not be an issue
if choosing between say, hot dogs and hamburgers, but with Labor/Leisure,
total output is limited by the number of hours in the day. The addition of
this constraint yields a maximization problem that is not generally
closed-form and convex. This can only amplify the behaviors we examine. We
stand by our labor/leisure narrative because it elegantly illustrates the
essential parts of our message. But for simplicity of optimization our math
ignores the constraint imposed by a 24-hour day.} Suppose the final producer
is a hot dog vendor choosing whether to provide snacks at a baseball game or
instead to take a seat and watch the Red Sox fight off their arch rivals
(who shall remain nameless.) The modification of the vendor's production
function or the cascading effects of modifications to intermediate
producers' production functions lead to unforeseen and sometimes highly
unpredictable and non-monotonic changes in the labor/leisure decision of the
vendor. In other words, unscalable individual decisions produce aggregate
outcomes that could not be easily modeled within the canonical neoclassical
framework.

\pgfplotsset{
    every axis/.append style={scale only axis, axis on top,
        height=4.25cm, width=6.25cm, xmin=-10, xmax=10, label style={font=\tiny}
        }
}

Figure \ref{funkyGraphs} illustrates four examples of the hot dog vendor
network. Relatively small changes to the producers' production functions
alter the output ratio considerably, and in highly unpredictable ways. Two
things are worth noticing. First, the hot dog vendor behaves almost
erratically as the input price changes. The ratio of Labor to Leisure takes
very different values in a relatively small input range. Second, small
changes in the Cobb-Douglas parameters yield substantial changes in the
shape of the graphs. This again underlines the difficulty of predicting
output variation caused by production changes in the hot dog technology.

Of course the choices of a single hot dog vendor are not enough to move the
needle on national unemployment or GDP figures. But the capacity of this
type of network to generate and transmit these erratic behaviors only
increases with the network's complexity. When Labor and Leisure choices are
writ large, unpredictable behavior can be catastrophic.

\begin{figure}[H]
\centering

\captionsetup{font=footnotesize,labelfont=bf}
\caption{The simple single-hidden-layer network architecture (inspired by
Rosenblatt''s perceptron) used in these functions is described in figure 
\protect\ref{networkDrawing}. See table \protect\ref{funkyGraphParams} for
initial parameters. \\[1em]
These hypothetical (but feasible) labor/leisure ratio functions are
nonmonotonic, even chaotic. As the log price of the single raw material
increases, the functions change regimes sharply and unpredictably. An
observer without a network-based economic model must describe such regime
changes as exogenous shocks. An additional source of such shocks is also
illustrated here: adjustments to the initial model parameters. These can
cause unexpected and counterintuitive changes to the labor/leisure ratio
function both in level and in shape. \\[1em]
The dashed line denotes the initial parameters, the dotted line (where
present) denotes intermediate parameters, and the solid line denotes the
final parameters. }
\label{funkyGraphs}
\end{figure}

\begin{longtable}[c]{cccccc}
\captionsetup{font=normal}
\caption{Parameters for Figure \ref{funkyGraphs} }\\\bottomrule\\
Model & Product Type & \(A^{Int/Leis}_i\) & \(\alpha^{Int/Leis}_i\) & \(A^{out}_l\) & \(\alpha^{out}_{lk}\)\\\midrule
Model 1 & Intermed & 9 & 0.675 & 1.684 & 0.091\\
 & " & 0.022 & 0.954 &  & 0.091\\
 & " & 0.389 & 0.964 &  & 0.13\\
 & " & 5,066 & 0.265 &  & 0.091\\
 & " & 2.2 & 0.53 &  & 0.01\\
 & Leisure & 11 & 0.53 & 1.5 & 0.01\\
 & " & 0.083 & 0.974 &  & 0.083\\
 & " & 2.399 & 0.964 &  & 0.091\\
 & " & 85.516 & 0.742 &  & 0.001\\
 & " & 22 & 0.53 &  & 0.142\\\midrule
Model 2 & Intermed & 0.002 & 0.909 & 1.6 & 0.375\\
 & " & 50 & 0.95 &  & 0.01\\
 & " & 0.383 & 0.98 &  & 0.057\\
 & Leisure & 0.406 & 0.968 & 1.5 & 0.091\\
 & " & 0.002 & 0.98 &  & 0.091\\\midrule
Model 3 & Intermed & 0.005 & 0.909 & 847.277 & 0.111\\
 & " & 41.389 & 0.909 &  & 0.067\\
 & " & 41.389 & 0.909 &  & 0.067\\
 & Leisure & 0.439 & 0.909 & 0.1 & 0.111\\
 & " & 34,600,299 & 0.909 &  & 0.067\\\midrule
Model 4 & Intermed & 47.275 & 0.476 & 6.009 & 0.01\\
 & " & 41.389 & 0.455 &  & 0.01\\
 & " & 2,105 & 0.417 &  & 0.01\\
 & " & 0.003 & 0.49 &  & 0.048\\
 & " & 0.541 & 0.484 &  & 0.038\\
 & Leisure & 34,600,299 & 0.455 & 1.5 & 0.01\\
 & " & 2.707 & 0.455 &  & 0.02\\
 & " & 0.002 & 0.455 &  & 0.231\\
 & " & 0.383 & 0.49 &  & 0.029\\
 & " & 5,530 & 0.476 &  & 0.005\\\bottomrule
 \\
 \captionsetup{font={footnotesize,stretch=1.2}}

\caption*{The network’s' initial parameters are described as follows: $%
A^{Int/Leis}_i$ refers to the Cobb Douglas technology parameter of the ith
intermediate or ith leisure good, $\alpha_i^{Int/Leis}$ refers to the CD
exponent on the sole (raw material) input of the ith intermediate or ith
leisure good. $A_l^{out}$ refers to the technology parameters of the output
producer, and $\alpha_{lk}^{out}$ refers to the output producer's exponent
on the kth input of type $l$  where $l \in$ \{Intermediate, Leisure\}.
}
\label{funkyGraphParams}
\end{longtable}

\section{A Simple Model of a Learning Economic Neural Network}

Imagine the fictional country of
Islandia, a small, isolated nation. Its
economy is driven by the import of raw materials and the export of finished
goods.

Islandia's imports consist of only two products: raw steel and brass. The
country uses these two raw materials to fashion intermediate goods which are
in turn used to fashion its single export commodity, clockwork chess players. Each
month, a cargo ship from the nearest Economic Super-Power offloads a
shipment of brass and steel and hauls away a load of Mechanical Turks to be
sold in catalogs, websites, and shopping malls. It takes exactly one month
for the small nation to convert brass and steel into chess players.

The producers in Islandia form a simple hierarchy. Eight work with the raw
steel and brass to create tools, tubing, spring steel, gear blanks, cog
blanks, and the like. These eight producers are the first level in the
economy, and their only inputs are the two imported raw materials. There are
then eight producers in the second level using some of each of the first
level's products as inputs. Levels three and four each have eight more
producers who use the previous level's outputs as inputs. The eight outputs
of level four are the sub-components used by the single level-five producer
to make clockwork chess players. So there are 33 individual producers in
Islandia, all of whom are organized into well-defined layers where 32
producers manufacture intermediate goods, and one producer manufactures the
final good.

\begin{figure}[h]
\centering 
\includegraphics[scale=0.4]{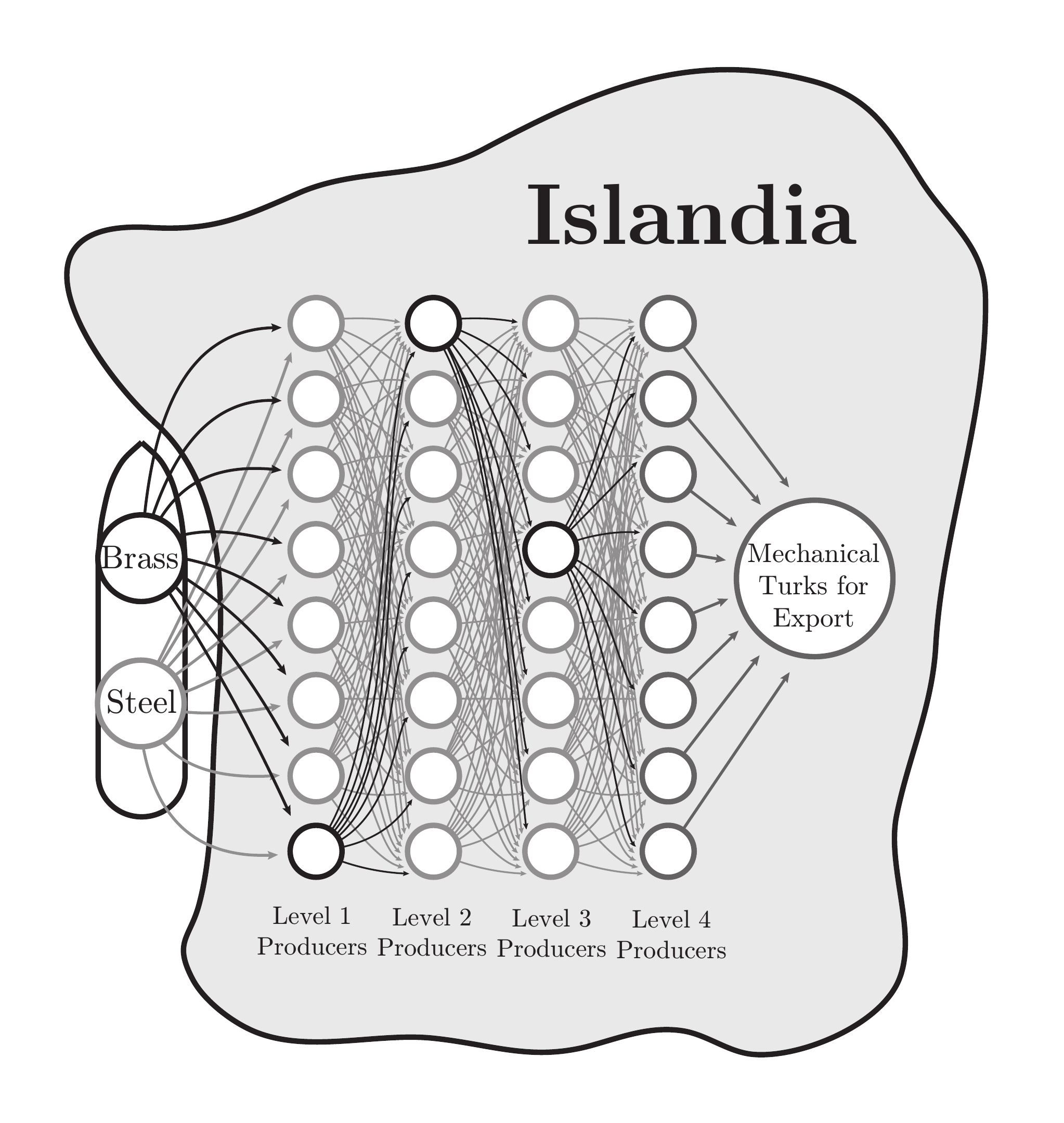} 
\caption{Input/Output matrix of Islandia economy. Arbitrary producers are
highlighted for visual clarity. \\[2em]
}
\end{figure}

The producers in Islandia all have Cobb-Douglas production functions of the
form:%
\begin{equation}\label{islandiaCD}
Y=A\prod_{i=1}^{n}x_{i}^{\alpha _{i}},
\end{equation}%
where $A$ is a technology parameter shared with all of the producers within
the same layer. The producers each attempt to maximize their own profits:%
\begin{equation}\label{islandiaProfit}
\pi =PY-\sum_{i=1}^{n}p_{i}x_{i}^{\alpha _{i}},
\end{equation}%
where $\mathit{P}$ is the anticipated price (the price the producer believes
she will get for her finished good at the time she makes her production
choices), $p_{i}$ is the price of the input commodity $x_{i}$, and $n$ is
the number of input commodities (always eight except for the first layer,
where it is two).

Prices are set through Walrasian tatonnement. Therefore, the accuracy of the
anticipated price cannot be known at the time of production. The actual
price manufacturer $m$ receives for her output minus the price she expected
to receive gives $m$ a value for her \textquotedblleft pricing
error.\textquotedblright\ She uses the pricing error to estimate her
production error -- that is, the amount that she over or under produced.
This production error could also be called, using the terminology of
Artificial Neural Networks (ANN's), her \textquotedblleft cost
function.\textquotedblright\ She is motivated to minimize this cost
function, and makes two types of adjustments to do so.

Once the producer has calculated her cost function, she first adjusts her
anticipated price. She uses a simple moving average for this adjustment.
Next, she employs a single step of the gradient descent procedure to make small adjustments to the exponents in her production function \citep{Ruder}. She is able to make these adjustments through slight
changes to the processes and technologies that she employs.

Let $E$ be the production error, or cost, let $\partial Y/\partial \alpha
_{i}$ be the partial derivative of output with respect to exponent $\alpha
_{i}$, and let $\mu $ be a \textquotedblleft learning
rate\textquotedblright\ parameter. Using gradient descent, producer $m$ will
update each of the exponents of her production function according to the
rule:%
\begin{equation}\label{updateRule}
\alpha _{i}^{\prime }=\alpha _{i}-E\mu \frac{\partial Y}{\partial \alpha _{i}%
}.
\end{equation}

All manufacturers on Islandia use the same learning process as producer $m$.
Their information set does not allow any of them to make adjustments to
their production based on the quantities of imports to or exports from the
nation -- but together, their economy forms an ENN that defines and refines decision boundaries -- regions of the
two-dimensional (brass and steel) import quantity map for which different
production levels are preferred.

The arrangement of the interconnections between neurons in a Neural Network
(NN) is known as the Neural Architecture \citep{Misra, Svozil}. Islandia has
a simple \textquotedblleft Feed Forward\textquotedblright\ architecture --
that is, each neuron outputs only to neurons on the next layer. There are no
outputs to neurons of the same or previous layers. Feed Forward networks can
vary by the number of \textquotedblleft hidden layers\textquotedblright\ (a
term referring to all layers except the input and output layers -- the Brass
and Steel and the Mechanical Turks in Islandia) and also the number of
neurons in each layer.

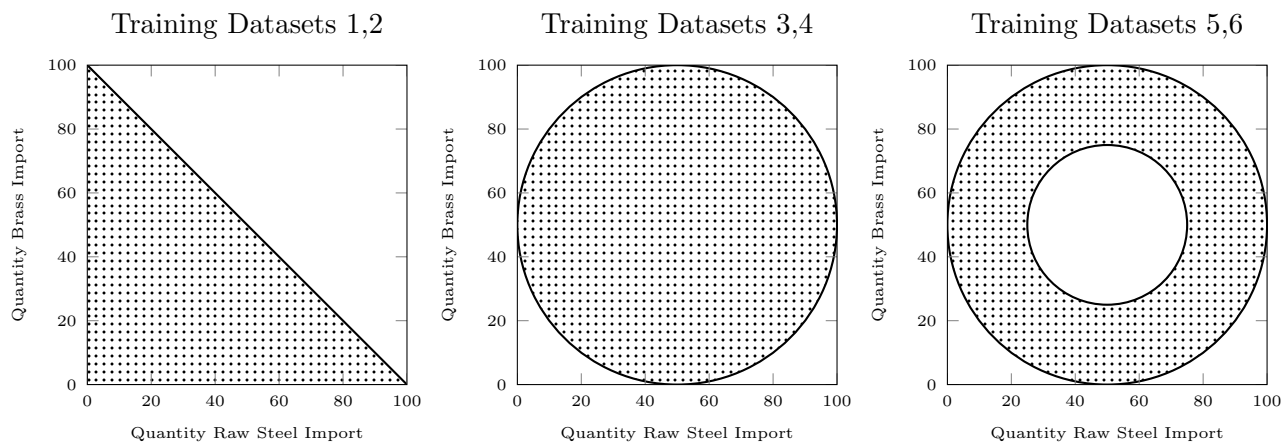
\begin{figure}[h]
\centering
\begin{tikzpicture}

\begin{axis}[ 
    name=plot1,
    title style={align=center},
    title={Training Datasets 1,2},
    ymin=0,
    xmin=0,
    xmax=100,
    axis equal image,
    xlabel={Quantity Raw Steel Import},
    ylabel={Quantity Brass Import}]
\addplot[black, thick][domain=0:100]{100-x};
\addplot [pattern=dots]coordinates {(0, 100)(100, 0)(0, 0)};

\end{axis}

\begin{axis}[ 
    name=plot2,
    title style={align=center},
    title={Training Datasets 3,4},
    at=(plot1.right of south east),
    anchor=left of south west,
    ymin=0,
    xmin=0,
    xmax=100,
    axis equal image,
    xlabel={Quantity Raw Steel Import},
    ylabel={Quantity Brass Import}]
\filldraw[color=black, pattern=dots, thick] (50,50) circle (50);
\end{axis}

\begin{axis}[ 
    name=plot3,
    title style={align=center},
    title={Training Datasets 5,6},
    at=(plot2.right of south east),
    anchor=left of south west,
    ymin=0,
    xmin=0,
    xmax=100,
    axis equal image,
    xlabel={Quantity Raw Steel Import},
    ylabel={Quantity Brass Import}]
\filldraw[color=black, pattern=dots, thick] (50,50) circle (50);
\filldraw[color=black, fill=white, thick] (50,50) circle (25);
\end{axis}

\end{tikzpicture}
\caption{ Maps of the six training dataset result values as a function of
input quantities. For odd numbered datasets the shaded region is true and
the unshaded region is false, vice-versa for even numbered datasets. Each
training dataset consists of 100 periods worth of randomly generated
input-result vector pairs. The 2D input vector consists of quantities
imported of steel and brass and the 1D boolean result vector is true if
demand in the period is higher than average. Using dataset 1 as an example,
let the randomly generated input vector be 40 units of steel and 40 units of
brass. The map defines the matching output vector as true. In contrast,
dataset 5 would map the same point to false. }
\label{decisionBoundaries}
\end{figure}

We note that the ENN on Islandia performs very poorly by Artificial Neural
Network standards. Unlike the NN's used in Artificial Intelligence
applications, the economy of the little nation is not optimized for network
performance. Every agent optimizes their choices based only on the local
price feedback mechanism (the local error), not on the component share in
the global error (as calculated by Islandia's export market and final
producer -- as would be the case in an efficient ANN.) Because price
information must trickle down the network layer by layer, the information a
producer uses to adjust production is delayed by several periods compared to
the information that was used to direct that production. This phase delay
scrambles much of the information about the economy's overall optimization
problem and prevents the country's ENN from learning to predict with
accuracies that rival ANNs. Economies more realistic than Islandia's are not as
highly constrained. Price is not the only information channel available to
most producers. Economic reports, stock market performance, communication
with suppliers and customers, and other business news sources give each
producer direct information about global system performance.

Another factor that prevents Islandia's ENN from approaching the performance
of an optimized ANN is that the Cobb Douglas production functions are
strictly increasing with respect to the input quantity of goods. As discussed above, neurons
within ANNs use functions that can be either increasing
or decreasing, as needed, with respect to their inputs. This means, for
example, that Islandia cannot manufacture more Mechanical Turks in response
to a shortage in both steel and brass -- even if the corresponding increase
in steel and brass prices consistently signals a boom in demand for clockwork
chess players next period. Although realistic ENN models can overcome this limitation, we keep it in place for Islandia in order to test learning behavior under the most restrictive assumptions.

All Neural Networks, whether ANNs, biological systems, or ENNs, are subject
to trouble with local optima. This is certainly an issue faced by each of
Islandia's producers and the country as a whole, but there is another
notable and idiosyncratic barrier to learning in Islandia. Every producer
has complete medium and long-term memory loss. Therefore
each producer can make exactly one adjustment to their production function
per period based only on the information in front of them. This is
equivalent to a hyper-restricted \textquotedblleft online
learning\textquotedblright\ paradigm in an ANN \citep{Misra, Svozil}.
High-power ANNs, however, typically use \textquotedblleft
batch-learning\textquotedblright\ processes. In batch-learning, the error
function is calculated from a large number of training examples (in
Islandia's case, periods) simultaneously. This is advantageous because a
single training example might lead a Neural Network to update in such a way
that it improves performance for that single example but degrades
performance for all others. Given the further possibility of local optima,
this means Islandia's online learning paradigm will sometimes lead to
network \textquotedblleft unlearning\textquotedblright\ -- that is, training
decreases performance rather than improves it. Fortunately, most real-world producers have a memory so this is only a problem in Islandia.  In summary, the Islandia model we test here is as suboptimal for learning as possible while still capable of testing whether prices are a sufficient information channel for online learning.  Our strategy is to isolate the price feedback mechanism as the source of training information while removing standard features of ANNs which would otherwise be strong assumptions in an economic model.  If the model exhibits global learning from price feedback under standard and weak assumptions, it is certainly capable of learning under stronger and more optimized assumptions.

\begin{longtable}[c]{c|ccc}
\captionsetup{font=normal}
\caption{Average Islandia ENN Learning Performance Over 20 Trials}\\\bottomrule\\
 & Initial Accuracy & Post-training Accuracy & Training Improvement\\
 & (Std. Error) & (Std. Error) & (t-score)\\[0.5em]\midrule\\
Dataset 1 & 20.15\% & 32.2\% & 12.05\%**\\
 & (0.69\%) & (3.35\%) & (3.3503)\\[0.5em]
Dataset 2 & 76.35\% & 90.3\% & 13.95\%***\\
 & (1.34\%) & (0.5624\%) & (9.586)\\[0.5em]
Dataset 3 & 70.2\% & 76.85\% & 6.65\%***\\
 & (0.73\%) & (1.0495\%) & (5.1905)\\[0.5em]
Dataset 4 & 27.65\% & 79.1\% & 51.45\%***\\
 & (0.84\%) & (0.764\%) & (45.2923)\\[0.5em]
Dataset 5 & 49.7\% & 54.8\% & 5.1\%*\\
 & (1.13\%) & (2.0424\%) & (2.1865)\\[0.5em]
Dataset 6 & 49.05\% & 54\% & 4.95\%**\\
 & (1.10\%) & (1.4654\%) & (2.7024)\\[0.5em]\bottomrule

\captionsetup{font={footnotesize,stretch=1.2}}

\caption*{Average Percent  Accuracy of Islandia's ENN before and after
training using the six different datasets illustrated in in Figure \ref{%
decisionBoundaries} on page \pageref{decisionBoundaries}. Each result is the
average of 20 independent trials with random initialization of the ENN's
parameters.  Training Improvement is defined as (Percent Accuracy
post-training) minus (Percent Accuracy pre-training).\\[0.5em]
*,**,and ***
denote greater than 95\%, 99\%, and 99.99\% confidence rejection of the
one-sided null hypothesis of no average training improvement.
}
\label{%
learningPerformance}
\end{longtable}

Even with Islandia's limitations its economy shows a remarkable
ability to learn.  Its output changes in response to patterns in
the input prices. Although this learning is driven by individual
microeconomic optimization choices, the individual producers are unaware
learning is occurring. It happens in the aggregate, on the macroeconomic
level, and not on the level of the producers who drive it.

Islandia is worth studying in this regard because of the simple structure of
its economy. We are able to create a precise and tractable computer model by
which to simulate its learning performance. The learning response of the
model of Islandia's economy with six different datasets is reported in Table %
\ref{learningPerformance}. The datasets consist of $100$ periods of import
supply and export demand data randomly generated to conform to the patterns
in Figure \ref{decisionBoundaries} on page \pageref{decisionBoundaries}. For
each period there is a pair of input quantities (representing raw steel and
brass), each ranging between $0$ and $100$, and a Boolean value assigned to
\textquotedblleft true\textquotedblright\ if Islandia can expect higher than
normal demand for its clockwork chess players. The six datasets each
exhibit a different pattern determining which regions of the input / output
map are true and which regions are false. The ENN attempts to converge to a
different optimal decision boundary for each dataset.

We initialized twenty independent model economies for each dataset, and in
each case trained the economies over $600$ randomly drawn rounds (each round
consisting of a single training example from the dataset.) To measure
learning, we first define the threshold output quantity as the average
output of the untrained economy across a subset of the training set inputs.
We then interpret an output above this threshold as corresponding to the
Boolean true value and an output below the threshold as corresponding to
Boolean false. The Percent Accuracy of the network is the proportion of
output true/false values that match the input true/false values. We define a
simple measure of learning performance as the increase in Percent Accuracy
-- that is, the Percent Accuracy of the trained network minus the Percent
Accuracy of the untrained network.

Learning performance is reported in Table \ref{learningPerformance} on page %
\pageref{learningPerformance}. Using dataset $1$, the ENN Percent Accuracy
increased\textbf{\ }$12.1\%$ on average. This was the most difficult dataset
for the ENN to respond to since lower input quantities of the raw materials
needed to correspond with higher output quantities of the finished good.
With dataset $2$, Percent Accuracy increased $14.0\%$ on average and with
dataset $3$ it increased $6.7\%$. However, for reasons that are the inverse
of the limitations on dataset $1$, the untrained ENNs performed very well on
these datasets from the outset -- both datasets averaging over $70\%$ before
training. Therefore there was less room for improvement from training.
Dataset $2$ yielded the highest average performance of all the datasets
after training, at $90.3\%$ Percent Accuracy. Dataset $4$ yielded an
accuracy increase of $51.5\%$. This was the dataset that elicited the
strongest learning response and the second highest absolute accuracy after
training, $79.1\%$. Datasets $5$ and $6$ have the most convoluted optimal
decision boundaries. They would likely benefit from a larger network since
the ability of a NN to converge to a complex and convoluted decision
boundary increases with the size of the network. Nevertheless, the ENN
exhibited learning behavior. Dataset $5$ accuracy increased by $5.1\%$ and
dataset $6$ accuracy increased by $5.0\%.$

\section{Concluding remarks}

Our macroeconomic understanding, until now, has been rooted in smooth
functions of aggregations of microeconomic agents. The possibility of
emergent economic phenomena has been implicitly recognized for some time,
but our assumptions may have inadvertently suppressed their study by
offering no mechanism for emergent behaviors to arise. We have analyzed two
critical examples of this. The endogeneity of apparently exogenous shocks
can be easily masked by the complexity of the ENN's network interactions.
Furthermore, because an ENN is capable of learning en masse, economic
structures (i.e. traditions, norms, or institutions) may evolve which
provide benefit to an aggregation without apparently providing utility to
the individual agents involved. Further study of these interactions may
yield novel insights into market failures like the Tragedy of the Commons, unfavorable game theoretical equilibria, and persistent informal structures such as race-based discrimination and inter-generational poverty.

Our studies of the ENN model are just beginning. Perhaps the most promising
implication of the ENN model is its use in government and NGO policy
formation. The ENN's learning behavior implies that policies might employ
active training techniques to affect change. We cannot say with confidence
that this is possible -- but the prospect is tantalizing enough to motivate
further study. It may be possible to identify a training `handle' -- by which
we mean an input through which to deliver rewards and punishments -- and a
low latency data source to observe an economic response to the varying
economic landscape.  In which case, regional and local economic reforms could be affected through training policies perhaps as simple as high frequency subsidy level adjustment.

We do not assert the ENN model supplants the canonical smooth models in any
way. These two models are complementary and coexist without conflict. Here
we appeal to the mathematical analogy between biological brains and ANN's,
which is quite robust. Medical science has learned much about the biological
brain, has employed powerful medical imaging, and has used effective
surgical and pharmacological techniques based solely on aggregate
approaches. Virtually no medical technology currently relies on the precise
identification of the synaptic weights of individual neurons. Aggregate
measures for diagnosis and prediction -- in both the biological brain and
the ENN -- will likely always be superior. Measurement precision and
computing power are finite. But the complexity of a Neural Network is vast
and perfect precision -- an impossibility -- would be required to accurately
predict its output. A simplified model cannot predict the output of a Neural
Network. Therefore, models of the ENN can only illuminate certain aspects of
its behavior. These are the topics of further research.





\bibliographystyle{econ}
\bibliography{RGBPbiblio}

\end{document}